# Propagation length of antiferromagnetic magnons governed by domain configurations

Andrew Ross[1,2,†], Romain Lebrun[1,†,*], Olena Gomonay[1], Daniel A. Grave[3], Asaf Kay[3], Lorenzo Baldrati[1], Sven Becker[1], Alireza Qaiumzadeh[4], Camilo Ulloa[5], Gerhard Jakob[1,2], Florian Kronast[6], Jairo Sinova[1,7], Rembert Duine[4,5,8], Arne Brataas[4], Avner Rothschild[3], Mathias Kläui[1,2,4,*]

[1]Institut für Physik, Johannes Gutenberg Universität-Mainz, 55099, Mainz, Germany.

[2]Graduate School of Excellence Materials Science in Mainz (MAINZ), Staudinger Weg 9, 55128, Mainz, Germany.

[3]Department of Materials Science and Engineering, Technion-Israel Institute of Technology, Haifa 32000, Israel.

[4]Center for Quantum Spintronics, Department of Physics, Norwegian University of Science and Technology, NO-7491 Trondheim, Norway.

[5]Institute for Theoretical Physics, Utrecht University, Princetonplein 5, 3584 CC Utrecht, The Netherlands.

[6]Helmholtz-Zentrum Berlin für Materialien und Energie, Albert-Einstein-Strasse 15, D-12489 Berlin, Germany.

[7]Institute of Physics ASCR, Cukrovarnicka 10, 162 53 Praha 6 Czech Republic.

[8]Department of Applied Physics, Eindhoven University of Technology, P.O. Box 513, 5600 MB Eindhoven, The Netherlands.

Keywords: Antiferromagnets, magnons, magnetic domains, XMLD-PEEM magnetic imaging, spin transport, magnon scattering

ABSTRACT

The compensated magnetic order and characteristic, terahertz frequencies of antiferromagnetic materials makes them promising candidates to develop a new class of robust, ultra-fast spintronic devices. The manipulation of antiferromagnetic spin-waves in thin films is anticipated to lead to new exotic phenomena such as spin-superfluidity, requiring an efficient propagation of spin-waves in thin films. However, the reported decay length in thin films has so far been limited to a few nanometers. In this work, we achieve efficient spin-wave propagation, over micrometer distances, in thin films of the insulating antiferromagnet hematite with large magnetic domains whilst evidencing much shorter attenuation lengths in multidomain thin films. Through transport and magnetic imaging, we conclude on the role of the magnetic domain structure and spin-wave scattering at domain walls to govern the transport. We manipulate the spin transport by tailoring the domain configuration through field cycle training. For the appropriate crystalline orientation, zero-field spin-transport is achieved across micrometers, as required for device integration.

MAIN TEXT

Recently, the focus of spintronics research has been shifting towards utilizing antiferromagnetic materials, with benefits over ferromagnets including terahertz spin dynamics and high stability against applied external magnetic fields, whilst the multiple sublattices of an antiferromagnet offer possibilities for novel phenomena not observed in ferromagnets.[1,2] Beyond reading[3,4] and switching[5,6] of the antiferromagnetic order, the manipulation of antiferromagnetic spin-waves could open pathways to both the development of ultrafast spin-logic devices, and to the realization of room temperature spin-superfluidity, the spin analogue to superconductivity.[7–9] However, the efficient transport of antiferromagnetic spin-waves has only been demonstrated in a mm thick antiferromagnetic single crystal,[10] whilst reports on thin film antiferromagnets have shown that spin-information only propagates across a few nanometers in both insulating[11–14] and metallic[15–18] antiferromagnets.

Spin-transport in antiferromagnetic insulators occurs through the excitation of the magnetic order, and the propagation of spin-waves, also known as magnons. Antiferromagnetic magnons can be either linearly- or circularly-polarized. However, only circularly-polarized magnon modes, present in antiferromagnets with an easy-axis anisotropy, carry angular momentum, permitting the transport of spin-information.[19] The spin-information of these magnons is carried (anti-)parallel to the equilibrium orientation of the antiferromagnetic Néel vector, $\boldsymbol{n}$.[10,20] The propagation of antiferromagnetic magnons can be monitored in an adjacent heavy metal layer acting as the source of detection and of excitation.[10,21] The generated signal strength is proportional to $\propto (\boldsymbol{n} \cdot \boldsymbol{\mu}_S)^2$, where $\boldsymbol{\mu}_s$ is the spin-accumulation in the heavy metal, leading to a maximal efficiency for $\boldsymbol{n}||\boldsymbol{\mu}_s$. The additional application of an external magnetic field, $H$, can be used to manipulate the direction of $n$ with respect to $\boldsymbol{\mu}_s$ and also serves to tune the frequencies of the contributing magnons. In particular, in an easy-axis (EA) antiferromagnet, $H||\hat{z}$(EA) leads to the well-known spin-flop transition, with the presence of a soft magnon mode[21,22] and by which the magnons undergo a transition from circular to linear polarization and the spin conductance is anticipated to diverge.[21]

The decay length of such excited magnons is expected to scale with their frequency,[15,23] and to depend on the magnetic damping of the material, which is reduced in insulating compounds due to the absence of conduction electrons.[24,25] This is in contradiction with the short propagation lengths reported in both metallic and insulating systems indicating that there is a mechanism in addition to the intrinsic magnetic damping dominating the transport of spin waves in this class of materials.

A particularly promising antiferromagnetic material for spin-transport is the iron oxide hematite, α-$Fe_2O_3$, an ubiquitous prototypical antiferromagnet commonly used in applications as diverse as humidity sensors[26] and photoelectrochemical cells.[27] Beyond these applications, it is also a promising material to investigate the propagation and manipulation of antiferromagnetic spin-waves using spintronic effects, given its low magnetic damping and intrinsic micrometer spin-diffusion length in bulk single crystals.[10] Hematite is also unique for its well-known Morin phase transition (MT),[28] where it undergoes a transition at a temperature $T_M$ from an easy-axis phase along the (0001) *c*-axis to an easy-plane phase where the magnetic moments lie perpendicular to (0001) ($T_M \sim$ 260 K in undoped bulk crystals). While long-distance spin-transport has been studied in bulk hematite,[10] investigating the transport in thin films will permit one to unravel the mechanism controlling the transport of magnons in thin film antiferromagnets. Answering this question could also permit one to realize the potential of this materials class in an applications-relevant thin film setting.

In this work, we address these open questions by realizing long-distance magnon transport in a thin film antiferromagnetic system and we identify the mechanisms governing the magnon propagation length by a systematic study. As the materials system, we employ epitaxial α-$Fe_2O_3$ thin films grown on sapphire substrates with different surface plane orientations.[29,30] The magnon propagation is investigated in these films as a function of the strength of a magnetic field applied along different directions with respect to the magnetic symmetry axes allowing for the extraction of the magnon propagation lengths of the different systems. To unravel the origin of the differences in the observed spin-transport length scales, the antiferromagnetic domain structure of the same films is determined by direct imaging and we identify, in conjunction with theoretical calculations, scattering at domain walls as the factor governing the propagation and thus the transport length.

First, we consider α-Fe$_2$O$_3$ grown along the (0001) direction. These films show a Morin transition at $T_M$ = 200 K which is hysteretic in nature[31] (details of the growth, Supporting Information), below which $\boldsymbol{n}$ is aligned along the $c$-axis, which is out of the plane as shown schematically in Fig. 1(a) (along $z$ for $T < T_M$).

When a charge current $I_{inj}$ flows through the Pt injector of a non-local structure (Fig. 1(a) and details in Supporting Information), a transverse spin current is created due to the spin Hall Effect (SHE) and a spin-bias $\boldsymbol{\mu}_s$ builds up at the Pt/α-Fe$_2$O$_3$ interface polarized within the sample plane along $\boldsymbol{y}$[32]. Given that a dissipation channel opens up as a spin flow across this interface, this spin-bias results in the excitation of spin-polarized magnons for a parallel alignment of the antiferromagnetic order and $\boldsymbol{\mu}_s$. A resulting non-equilibrium magnon population builds up in the α-Fe$_2$O$_3$ for $\boldsymbol{\mu}_s \| \boldsymbol{n}$ that then diffuses away from the injector.[10,33] The polarization of the non-equilibrium magnons, whether linear or circular, will depend on the symmetries of the antiferromagnet being investigated. The magnonic spin current flowing in the antiferromagnet is then absorbed by a Pt detector due to the non-equilibrium between the magnon population of the α-Fe$_2$O$_3$ and the electron population of the Pt at the interface,[10,33] whereby it is converted to a charge current via the inverse SHE. This spin-bias signal can then be expressed as a non-local resistance ($R_{el} = V_{el}/I_{inj}$).

We measured $R_{el}$ for selected different antiferromagnetic configurations with the magnetic fields in the sample plane, parallel to the wires (which are parallel to *x*) and out of the plane, along the easy-axis (along *z*), Fig. 1. Here, we focus on measurements performed at 175 K, below $T_M$, in the easy-axis phase where we can stabilize and propagate circularly-polarized magnons.

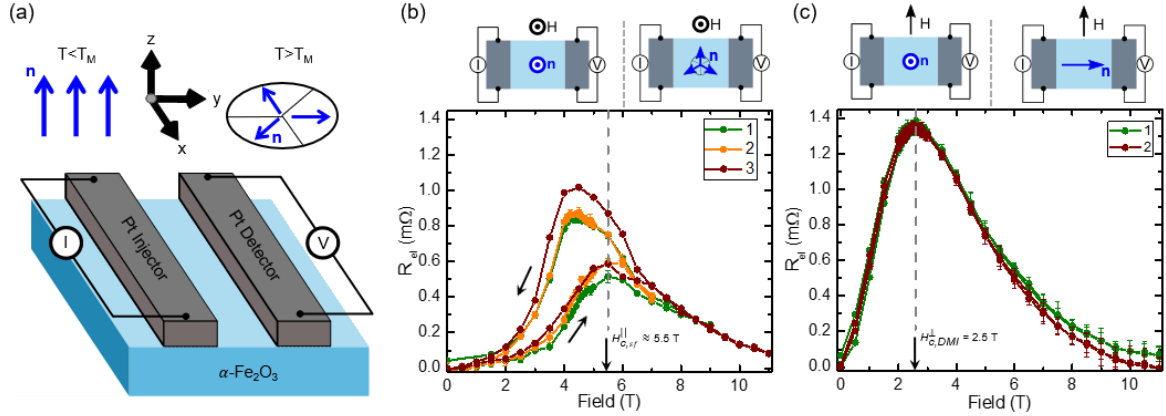

**Figure. 1.** (a) Schematic of the used non-local geometry of platinum wires on top of an antiferromagnetic insulator thin film. The electrical contacts and axes are indicated as well as the relative orientation of the Néel vector (blue) in respectively the easy-axis ($T<T_M$) and easy-plane ($T>T_M$) phases (here assuming an additional threefold in-plane anisotropy) for the 100 nm thick (0001) films. (b) Spin-bias signal for **H**||**z** and a wire separation of 750 nm in geometry two, scaled appropriately. The spin-bias signal displays a hysteretic-like behavior due to an additional magnetic anisotropy of magnetoelastic origin above $\mu_0 H_{c,sf}^{\parallel} \approx 5.5$ T that needs to be overcome when lowering the field and shows a training effect with field cycling (1-3) that is reset when warmed above $T_M$. (c) Spin-bias signal for **H**||**x** and a wire separation of 750 nm wires in geometry two, scaled appropriately. The spin-bias signal increases up to $\mu_0 H_{c,DMI}^{\perp} = 2.5$ T, where the Néel vector ***n*** is aligned along **y**. This orientation shows no training effect or hysteresis.

For a field applied along *y*, perpendicular to the non-local Pt wires, as well as at zero field, we observe no spin-bias signal arising from the injected spins as in all these cases $\boldsymbol{n} \perp \boldsymbol{\mu}_s$. On the contrary, maximal signals for $R_{el}$ are observed for magnetic fields perpendicular to the wires in ferromagnetic materials.[33] For a field $\mathbf{H} \| \boldsymbol{z}$ (Fig. 1(b)), we observe the spin-flop transition at around $\mu_0 H_{c,sf}^{\|} \approx 5$ T, above which $\boldsymbol{n}$ lies in the sample plane (see Fig. 1(a)). The transmitted signal reaches a maximum in close vicinity to $H_{c,sf}^{\|}$. As the field is reduced back to zero, we see a hysteresis like behavior emerging, where the $R_{el}$ is larger for some field range than for the increasing field branch. By repeated field cycling, we also see a clear training effect, where the peak amplitude increases and the transition width becomes sharper with successive cycles indicating changes in the magnetic properties. As a first key result, this demonstrates that the properties of the spin transport are not given just by the magnetic system but can be additionally tuned by magnetic field effects.

We ascribe the observed enhancement of the transport with applied field to the strong induced softening of one of the magnon modes[10,22,34] in the vicinity of spin-flop transition and the corresponding increase of the magnon population as the magnon gap is reduced. Below $H_{c,sf}^{\|}$, $\boldsymbol{n} \perp \boldsymbol{\mu}_s$ and the signal is suppressed whilst in the transition region around $H_{c,sf}^{\|}$, $\boldsymbol{n}$ has a non-zero projection on $\boldsymbol{\mu}_s$ leading to a pronounced contribution to the signal. As the field is further increased ($H > H_{c,sf}^{\|}$), the excited magnons adopt a linear polarization in the easy-plane phase, the magnon gap opens and the signal diminishes.

The hysteretic behavior observed, along with the training effect in these (0001) thin films is distinctly different from the behavior found previously for single crystals.[10] To understand this behavior, we need to firstly take into account the particular geometry where the field is applied along the easy-axis perpendicular to the surface, which has not been previously measured for single crystals. In this case above the spin-flop with **H**∥*z*, ***n*** is in the easy-plane phase, which exhibits a 3-fold in-plane degeneracy. This leads to the formation of domains where, due to this 3-fold degeneracy, the Néel vector rotates by 60° in adjacent domains. Successive field cycles through the spin-flop transition lead to depinning and annihilation of domains walls, and given the pronounced increase of the signal this is a first indication that the spin-transport depends on the domain configuration. The domain formation above the spin-flop is dominated by the magnetostrictive interaction, which can be sizeable in hematite,[35] and additionally imperfections of the substrate as the growth seed can have an effect. Such an additional magnetostrictive-induced anisotropy above $H_{c,sf}^{\parallel}$ leads to a pinning potential that must be overcome when the magnetic field is lowered thus leading to the hysteretic signal due to hysteretic changes in the domain structure. From a hysteresis width of ~ 1.5 T, we calculate a uniaxial anisotropy field of 30 mT and magnetostrictive field of 8 mT. From this, we can estimate the strength of the pinning to be $8 \times 10^2 \frac{J}{m^3}$ in agreement with literature values on bulk samples[35] (details, see model in Supporting Information).

Beyond this first geometry of the applied field direction, we next apply the field along the wires, **H**∥*x* (see Fig. 1(c)). We also observe a spin-bias signal in this geometry. We note that in easy-axis antiferromagnets that also have a bulk Dzyaloshinskii-Moriya interaction (DMI) parallel to the easy-axis, a spin-reorientation for ***H*** ⊥ **EA** can occur.[36-38] This transition is not seen in simple easy-axis antiferromagnets without bulk DMI such as $Cr_2O_3$ and is a smooth second-order transition, unlike the hysteretic spin-flop transition at $H_{c,sf}^{\parallel}$. In an applied field perpendicular to the EA, the spins begin to cant towards the direction of *H*, leading to an arising perpendicular component of the DMI. This component then acts to align the spins perpendicular to the easy-axis given its antisymmetric nature. The net result is such that, from zero field to a critical field $H_{c,DMI}^{\perp}$, the spins smoothly rotate and align perpendicular to both the external field and the DMI vector (∥EA). This leads to a progressive parallel alignment of *n* and *μ*s, accompanied by the softening of one magnon mode[37,38] allowing us to draw strong parallels to magnetic field transitions along respectively the hard and easy-axis of a ferromagnet. Similarly to the **H**∥**z** case, these two features explain the increasing signal up to $H_{c,DMI}^{\perp}$ where *n*∥*μ*s. The signal is furthermore exhibiting a π periodicity of the angle between *n* and *μ*s (see Supporting Information). Above this critical field of the spin-reorientation, the signal again decreases at large fields due to the emergent linearly-polarized magnons and an opening of the magnon gap.[37,38] In contrast to the case **H**∥**z**, no additional domain structure is formed when **H**∥**x**, as the equilibrium orientation of *n* is nondegenerate, leading to the absence of any hysteresis or training effects. We note here that these measurements were performed 25 K below the measured Morin transitions of these films, however the same signals are observed in the clean easy-axis phase at 100 K, whilst the signal is drastically different at 290 K in the clean easy-plane phase (see Supporting Information).

As a magnonic spin current is transmitted when *μ*s and *n* are parallel, in the (0001) oriented films this requires the application of high magnetic fields. Such fields are impractical and inefficient for applications that make use of antiferromagnetic components, in addition to the problem of the ambiguity of the signal in the "flopped" state for H∥EA as the magnon modes transition from circularly to linearly polarized. It has also been shown that in ferromagnetic systems, the application of large magnetic fields strongly modifies the spin diffusion length[39] although it is unclear if such an effect is also present in antiferromagnets.[15] Therefore, rather than using magnetic fields to align *n* and *μ*s, a zero-field projection would not only be more practical for antiferromagnetic transport applications, but also allow for the investigation of the transport properties of antiferromagnetic magnons without the modification of the magnon spectra.

Therefore we next grow ($1\bar{1}02$) (r-plane) oriented thin films of hematite which exhibit a Morin-temperature $T_M$=200 K (see Supporting Information), and have both a stable surface termination[40] and a zero-field projection of *n* onto the surface plane. For wires perpendicular to the EA projection (Fig. 2(a)), a parallel alignment of *n* and $\mu_s$ is possible below the spin-reorientation. Thus, we can study potential zero-field transport as a function of external field perpendicular (**H**||*x*, Fig. 2(b)) and parallel (**H**||*y*, Fig. 2(c)) to the wires. A significant signal is observed even at zero field where the magnon polarizations have a nonzero projection on $\mu_s$. With increasing field the signal increases until *n* rotates perpendicular to $H$[41] and the field dependence shows no training effect or hysteresis. The suppression of the signal with increasing field above the spin-flop transition has the same origin as for the previously discussed (0001) growth orientation. This crystallographic orientation thus highlights that an efficient magnonic transport is possible through a careful choice of geometry, even without an applied field.

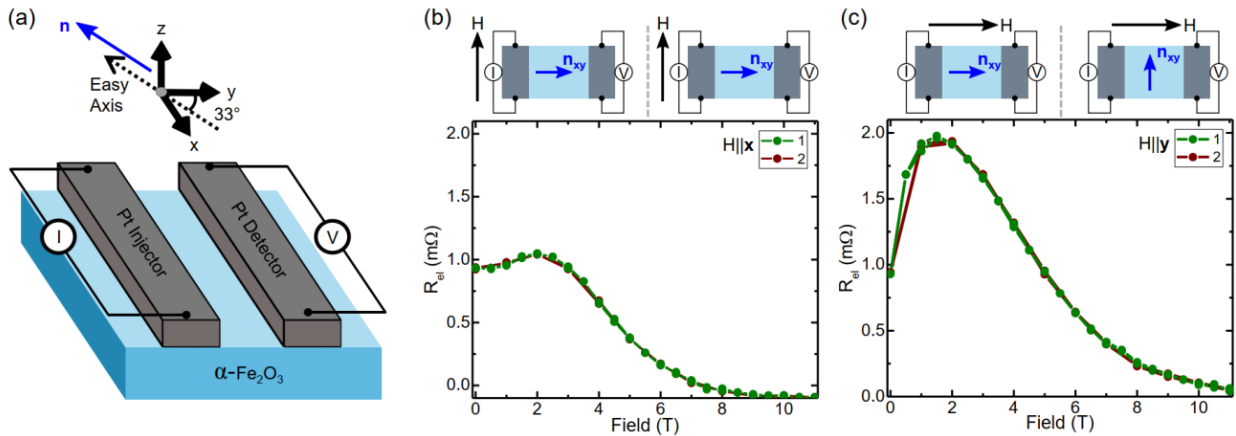

**Figure 2.** (a) Sketch of non-local geometry with respect to the magnetic axis for 500 nm thick ($1\bar{1}02$) hematite films. The **EA** is tilted 33° in the *yz* plane, indicated by the dashed line. The Néel vector lies parallel to the **EA** at zero-field. (b, c) Spin-bias signal for ($1\bar{1}02$) hematite with wires

separated by 850 nm perpendicular to the projection of the **EA**. The field is directed along (b) **H**||**x** and (c) **H**||**y** for two cycles (1,2). There exists a significant spin-bias signal at zero-field that is suppressed at high fields as the Néel vector rotates. There is no observed training effect or hysteresis for this geometry. The error bars are the standard deviation of the data points.

The key question is then the propagation length, and whether information can be transported across sufficient distances to allow for the development of spin-logic devices. While previously transport in thin film antiferromagnets has only been possible over few nm distances, we can transport spins here over µm as shown in Fig. 3 (the value of $R_{el}$ was recorded at both $H_{c,sf}^{||}$ and $H_{c,DMI}^{\perp}$ in the (0001) films whilst for the (1$\bar{1}$02) films, the value at $\mu_0 H$ = 1.5 T for **H**||**x** and **H**||**y** is taken to remove any possible interfacial canting [42]). The lack of a threshold[7,8] and the used elevated temperature of 175 K indicate that diffusive rather than superfluid transport dominates,[10] allowing us to use a one-dimensional spin-diffusion-relaxation equation to estimate the spin-diffusion lengths in our films.[33] We find for (0001) and (1$\bar{1}$02) samples for the three selected geometries: $\lambda_{c,DMI}^{\perp}$ = 350 nm, $\lambda_{r,1.5\,T}^{x}$ = 600 nm, and $\lambda_{r,1.5\,T}^{y}$ = 800 nm, which is two orders of magnitude greater than previous reports for epitaxial[11] or polycrystalline[12] thin film antiferromagnets. However, one must also note that the distance dependence of $R_{el}$ at $H_{c,sf}^{||}$ cannot be fitted with a simple 1D spin-diffusion model and, for this situation, the hysteretic behavior and the observed training effects do not allow for an unambiguous determination (inset of Fig. 3). Together with the observation that the signal during the hysteretic behavior depends on the field cycling which affects the spin structure, this raises the major open question of the origin of the transport of magnons across distances far larger than previously observed in thin film antiferromagnets. Furthermore, we need to explain why the spin-transport length-scales vary so strongly even within a single material and cannot be described for all geometries by a simple diffusion model.[33]

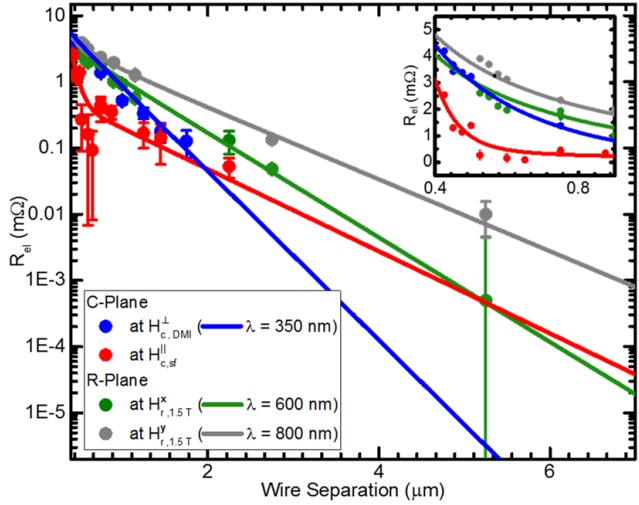

**Figure 3.** Distance dependence of non-local resistance $R_{el}$ for (0001) hematite at both $H_{c,sf}^{\parallel}$ (red) and $H_{c,DMI}^{\perp}$ (blue) as well as $(1\bar{1}02)$ hematite for a field directed along $\boldsymbol{x}$ (green) and $\boldsymbol{y}$ (grey) along with fits to the one-dimensional spin-diffusion equation. The solid red line of $H_{c,sf}^{\parallel}$ represents a diffusion equation with two characteristic lengths. Inset shows the spin-bias signal for low distances. The error bars represent the standard deviation of the data. For $H_{c,sf}^{\parallel}$ the error bars represent the standard deviation of multiple field cycles. The curve at $H_{c,sf}^{\parallel}$ (red) is the average across many successive field cycles for each device at the peak field for the lowering field sweep.

To analyze the mechanisms that govern the propagation lengths, we consider two factors. The decay of magnons due to magnetic damping leads to an attenuation that is, in the simplest model, a materials property and independent of the spin structure. Additionally, scattering of magnons is predicted to occur at magnetic domain walls[43,44] leading to a spin-structure dependent propagation length and could serve as a possible explanation for the short diffusion lengths previously seen. To investigate these factors, we performed X-ray magnetic linear dichroism photoelectron emission (XMLD-PEEM) imaging at the Fe L2 edge (details, see Supporting information). Whilst the domain structure of the ($1\bar{1}02$) films at zero-field can be imaged directly, the domain structure in the (0001) films at either $H_{c,sf}^{\parallel}$ and $H_{c,DMI}^{\perp}$ cannot be directly observed due to the need for high magnetic fields that cannot be applied during PEEM imaging. However, the type of domain structure at the spin-flop field can be obtained by making use of the temperature-field equivalence for hematite and instead imaging the spin-structure above $T_M$, where the magnetic state is also in the easy-plane and energetically a similar situation is attained.[36] For the (0001) films, we observe in Fig. 4(a)-4(c) (100 K – 300 K) a multi-domain pattern. Above $T_M$ (Fig. 4(c)) we find domain sizes of a few hundred nanometers. When we reduce the temperature, the domain spin structure changes strongly across the Morin transition to a state that is reproducible when cooling. Surprisingly, we still measure a contrast from magnetic domains below $T_M$ (see Fig. 4(a)) where the spin-structure should be in the easy-axis phase, and 180° domains should lead to the same dichroic signal and the absence of contrast. We associate this contrast to a distribution of the growth crystallites leading to an off-axis component of *n* (see Supporting Information). However, we can infer the presence of 180° domain walls below $T_M$ from the clear observation in the ($1\bar{1}02$) films (see Fig. 4(d)) as discussed later.[45] Whilst the domain structure above $T_M$ gives an indication of the magnetic state at $H_{c,sf}^{\parallel}$, no direct, quantitative way exists to access the magnetic state at $H_{c,DMI}^{\perp}$. Nevertheless, the domains are expected to be similarly sized and lack degeneracy at $H_{c,DMI}^{\perp}$, as evidenced by the non-hysteretic transport signal (Fig.1(c)), separated by 180° domain walls.

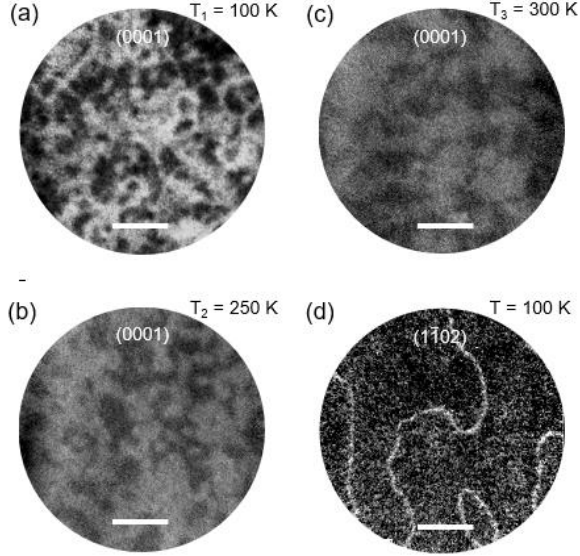

**Figure 4.** (a)-(c) XMLD-PEEM images of 100 nm thick (0001) hematite at $T_1 = 100$ K, $T_2 = 250$ K and $T_3 = 300$ K, and $\mu_0 H = 0$ T. (d) XMLD-PEEM imaging of 500 nm (1$\bar{1}$02) hematite at $T = 100$ K, below the Morin transition, and $\mu_0 H = 0$ T. Large, micrometer sized domains are separated by clear 180° domain walls. The scale bars represent 2 μm.

In the (1$\bar{1}$02) films, we observe below $T_M$ and at zero-field that 180° domain walls separate large single domain areas and the contrast is larger than for the (0001) orientation due to the inclination of the easy-axis (see Fig. 4(d)). To our knowledge, the clear observation of 180° antiferromagnetic domain walls in antiferromagnetic films has not been reported, and so far only 90° domain walls have been observed.[46] The size of the domains is of the order of many micrometers, an order of magnitude larger than in the (0001) orientation. The (1$\bar{1}$02) films thus show not only longer propagation lengths than (0001) films but also larger antiferromagnetic domains, highlighting the relation between the domain structure and the spin-transport.

To understand this behavior, we discuss the propagating magnonic spin current properties and its interaction with domains and domain walls. The magnons are excited with a broad spectrum of energy and wavevector (***k***) with an upper limit determined by the thermal energy. Within a single domain, the magnon propagation length is mainly determined by spatial diffusion with a length scale governed by the magnetic damping. However, the presence of domain walls introduces additional dissipation channels through spin-dependent scattering.[43,44,47] This can reduce the propagation length of the magnons and experimentally lead to an attenuation of the signal. At 60° domain walls, such as above $H_{c,sf}^{\parallel}$, magnons scatter due to the Néel order reorientation between the neighboring domains, however the question of the behavior at a 180° domain wall remains.

For circularly-polarized magnons, as in the easy-axis phase of hematite, we now develop a simple toy model of the scattering process at a 180° domain wall pinned by a homogeneous distribution of defects acting to fix the orientation of ***n*** in the center of the domain wall. The dynamics of the space (**r**) distribution of the Néel vector ***n*(r, *t*)** (|***n***|=1) in an antiferromagnetic texture is described within the Lagrange formalism.[48,49] By solving the Lagrange function, we can establish the expression of a pinned domain wall (including its width $x_{DW}$) in the presence of uniaxial anisotropy and homogeneous pinning around the domain wall (see Supporting Information). The magnons in the domain wall are the perturbations of the Néel vector on top of the static magnetic texture. In the limit of a pinning width much smaller than the domain wall width ($x_{def} \ll x_{DW}$), the domain wall leads to the reflection of incident plane waves with a reflection coefficient given by,

$$R = \frac{4(E_{\text{pin}}/E_{\text{DW}})^2}{4(E_{\text{pin}}/E_{\text{DW}})^2 + k^2 x_{\text{DW}}^2}, \quad (1)$$

with $x_{DW}$ the domain wall width, $E_{\text{pin}}$ and $E_{\text{DW}}$ respectively the pinning and domain wall energies. As seen from this equation, pronounced reflection (with the reflection coefficient R ≤ 1) occurs for magnons with a wave-vector $k \leq k_{crit}$, where the critical value is the upper limit of the reflecting potential given by,

$$k_{\text{crit}} = \frac{4E_{\text{pin}}}{E_{\text{DW}}} \frac{1}{x_{\text{DW}}}, \qquad (2)$$

which depends on the ratio between the pinning and the domain wall energies.

It should be mentioned that although equation (1) is derived under the assumption that $H=0$, it is also valid for $H \leq H_{sf}$, the spin flop field, where the energy of the domain wall now adopts the field-dependence given by, $E_{\text{DW}} \rightarrow E_{\text{DW}}\sqrt{1 - H^2/H_{sf}^2}$. This means that in the vicinity of the spin-flop field, the reflection of magnons substantially increases $k_{\text{crit}} \gg 1/x_{\text{DW}}$.

From this model, we find that the scattering is significant up to a critical value of the wavevector, $k_{crit}$, which is proportional to the pinning field obtained from the experimental data. For large pinning, all types of antiferromagnetic magnons can scatter from the domain walls and the propagation length is dominated by the domain size.

This now allows us to explain the observed transport: with magnetic domains larger than the diffusion length, the magnon transport obeys a 1D diffusive model in the $(1\bar{1}02)$ films. In in the $(0001)$ films, with the presence of many domain walls that act as scattering sites, the signal attenuates over shorter distances where the size of the domains serve to define the characteristic decay length. This additional scattering can then be described by a set of diffusion equations,

$$D_k \Delta \rho_k - \frac{1}{\tau}\rho_k = 0, \qquad (3)$$

where the diffusion coefficients $D_k$ depend on the magnon wavenumber $k$, $\rho_k$ is the number of magnons with the certain $k$ per unit volume, $\tau$ is the relaxation time whose dependence on $k$ can be neglected. As the domain walls are effectively transparent for magnons with a large wavevector, $k > k_{crit}$, the corresponding diffusion coefficient is that of a single domain sample, $D_k = D_0$.

However, for magnons with a wavevector less than $k_{crit}$, the scattering from the domain walls greatly reduces the effective rate of diffusion. For simplicity, we then assume that in this case, $D_k \sim D_1$ for all magnons below this threshold, $k<k_{crit}$. Within this simplified picture, the electrical signal, i.e. $R_{el}$, measured at a distance $x$ is then proportional to the total magnon flux,

$$R_{el}(x) \propto \sum_k \rho'_k(x) \approx \rho_0 \left( \frac{f}{\lambda_1} \exp\left(-\frac{x}{\lambda_1}\right) + \frac{1-f}{\lambda_0} \exp\left(-\frac{x}{\lambda_0}\right) \right) \quad (4)$$

where $\rho_0$ is the total number of magnons created in the vicinity of the injector due to the spin accumulation $\mu_s$, $f$ is the fraction of magnons with $k<k_{crit}$ and $\lambda_{0,1} = \sqrt{D_{0,1}\tau}$ is the diffusion length of the magnons with respectively $k<k_{crit}$ and $k>k_{crit}$. We use equation (4) to fit the experimental data in Fig. 3 for the value of $R_{el}$ in the (0001) film at $H_{c,sf}^{\parallel}$ with $\lambda_1 = 67$ nm and $\lambda_0 = 700$ nm. We present in the Supporting Information a summary of the diffusion lengths for the curves in Fig. 3 alongside the size of the domains imaged in Fig. 4 extracted from intensity profiles of the PEEM images.

We thus find that magnons with a high wavenumber, in other words the high frequency magnons that one would aim to use for antiferromagnetic devices, diffuse with a characteristic length comparable to the other geometries discussed so far whilst the low frequency magnons are scattered with a secondary characteristic length scale qualitatively governed by the domain structure (Fig. 4). There has been increasing interest in the past few years in the transport regime of magnons in antiferromagnets, including in the diffusive[50] and in particular the superfluid regimes.[8,51] Such a possible magnonic supercurrent can exist in antiferromagnets with an easy plane anisotropy, such as above $H_{c,sf}^{\parallel}$, however, it suffers from phase slips that act to disrupt the precession of *n* within the easy-plane.[52,53] In the presence of spin structures like domain walls, the flowing magnons can be absorbed by the spin texture[54] or they may even act as conduits for the flowing spin current.[55] However, the orientation of the domain walls, and thus the rotation of ***n*** across them, is both parallel and perpendicular to the direction of propagation, which would hinder any possibility for spin transport in the superfluid regime. The small domains frequently observed in thin film antiferromagnets[3,45,46] thus prove detrimental to the transport of spin information, regardless of transport regime when the excited magnons are lower in frequency such as those injected by ferromagnetic layers at resonance[12] or from broadband excitations such as performed here and in other studies[11,15] where the scattering acts as a filter on the propagating magnon current. We here highlight that the functionalization of antiferromagnetic spintronic devices strongly benefits from either the absence of antiferromagnetic domain walls or from higher frequency excitations.

To conclude, the transport of spin-information in antiferromagnetic thin films with large magnetic domains can reach micrometers, as in the best ferromagnetic thin film systems,[33] which opens the way for instance towards the development of antiferromagnetic spin-logic devices.[56] An applied magnetic field can be used to both train the domain structure and control the orientation of the Néel vector in order to facilitate the detection and tune the transport efficiency of magnonic spin transport. For selected growth orientations, a zero-field projection of the Néel vector can be achieved leading to field free spin-transport. Thus our identified different approaches to control magnon transports offer promising routes to the application of antiferromagnets in magnonic devices.[56]

**Corresponding Author**

*rolebrun@uni-mainz.de. *klaeui@uni-mainz.de.

**Author Contributions**

R.L. and M.K. proposed and supervised the project. A.R. performed with R.L. the transport experiments. R.L., L.B. and F.K. performed the XMLD-PEEM measurements and analyzed the data with A.R. A.R. patterned with R.L. the samples. D.A.G., A.K. and Av.R. grew and optimized the films. A.R., R.L. and S.B. performed the XRD and SQUID measurements with input from G.J. O.G., C.U. and A.Q. developed theoretical understanding with assistance and input from J.S., R.D. and A.B. A.R. and R.L. analyzed the data with inputs from M.K. and O.G. A.R. and R.L. wrote the paper with M.K. and O.G. All authors commented on the manuscript. *These authors contributed equally.

ACKNOWLEDGEMENTS

The authors acknowledge thoughtful input and discussion with David Ellis. We also thank Helmholtz Zentrum Berlin for the allocation of synchrotron radiation beamtime. A.R., G.J., and M.K. acknowledge support from the Graduate School of Excellence Materials Science in Mainz (DFG/GSC 266). A.R., R.L. and M.K. acknowledge support from the DFG project number 423441604. R.L. acknowledges the European Union's Horizon 2020 research and innovation programme under the Marie Skłodowska-Curie grant agreement FAST number 752195. S.B. and G.J. from DFG project number 358671374. O.G. and J.S. acknowledge the support from the Humboldt Foundation, the ERC Synergy Grant SC2 (No. 610115), the EU FET Open RIA Grant no. 766566. O.G. additionally acknowledges the DFG (project SHARP 397322108). D. A.G., A.K. and Av.R. acknowledges support from the European Research Council under the European Union's Seventh Framework programme (FP/200702013) / ERC (Grant Agreement No. 617516). D.A.G. acknowledges support from The Center for Absorption in Science, Ministry of Immigrant Absorption, State of Israel. L.B. acknowledges the European Union's Horizon 2020 research and innovation programme under the Marie Skłodowska-Curie grant agreement ARTES number 793159. All authors from Mainz also acknowledge support from both MaHoJeRo (DAAD Spintronics network, project number 57334897) and SPIN+X (DFG SFB TRR 173, projects A01 and B02). A.Q. and A.B. acknowledge support from the European Research Council via Advanced Grant number 669442 'Insulatronics'. A.Q., C. U., R.A.D., M.K. and A.B. were supported by the Research Council of Norway through its Centres of Excellence funding scheme, project number 262633 'QuSpin'.

Supporting Information

The supporting information contains details of the growth of the hematite films as well as descriptions of the structural and magnetic characterization. The measurement procedure for the transport measurements is also presented along with the angular dependence of the signal for a fixed magnetic field. Next, the signals for the two field directions in the (0001) films is shown at temperatures far above the Morin transition and far below. Finally, details and derivation of the model used to describe the scattering of magnons from 180° antiferromagnetic domain walls is included.